\documentstyle[twocolumn,aps,prl,floats,psfig]{revtex}
\begin{document}
\draft

\title{Precision Measurement of the Spin-dependent Asymmetry \\ 
       in the Threshold Region of $^3\vec{\mathrm{He}}(\vec{e},e')$}
 
\author{F.~Xiong,$^{12}$ D.~Dutta,$^{12}$ W.~Xu,$^{12}$ B.~Anderson,$^{10}$ 
L.~Auberbach,$^{19}$ 
T.~Averett,$^{3}$ W.~Bertozzi,$^{12}$ T.~Black,$^{12}$ J.~Calarco,$^{22}$ 
L.~Cardman,$^{20}$ G.~D.~Cates,$^{15}$ Z.~W.~Chai,$^{12}$
J.~P.~Chen,$^{20}$ S.~Choi,$^{19}$ E.~Chudakov,$^{20}$ 
S.~Churchwell,$^{4}$ G.~S.~Corrado,$^{15}$ C.~Crawford,$^{12}$ 
D.~Dale,$^{21}$ 
A.~Deur,$^{11,20}$ P.~Djawotho,$^{3}$ B.~W.~Filippone,$^{1}$ 
J.~M.~Finn,$^{3}$ 
H.~Gao,$^{12}$ R.~Gilman,$^{17,20}$ A.~V.~Glamazdin,$^{9}$ 
C.~Glashausser,$^{17}$ 
W.~Gl\"{o}ckle,$^{16}$ J.~Golak,$^{16,8}$
J.~Gomez,$^{20}$ V.~G.~Gorbenko,$^{9}$ J.-O.~Hansen,$^{20}$ 
F.~W.~Hersman,$^{22}$ D.~W.~Higinbotham,$^{24}$ R.~Holmes,$^{18}$ 
C.~R.~Howell,$^{4}$ E.~Hughes,$^{1}$ B.~Humensky,$^{15}$ S.~Incerti,$^{19}$ 
C.W.~de Jager,$^{20}$ J.~S.~Jensen,$^{1}$ X.~Jiang,$^{17}$ 
C.~E.~Jones,$^{1}$ M.~Jones,$^{3}$ R.~Kahl,$^{18}$ H.~Kamada,$^{16}$ 
A. Kievsky,$^{5}$ I.~Kominis,$^{15}$ W.~Korsch,$^{21}$
K.~Kramer,$^{3}$ G.~Kumbartzki,$^{17}$ M.~Kuss,$^{20}$ 
E.~Lakuriqi,$^{19}$ M.~Liang,$^{20}$ N.~Liyanage,$^{20}$ J.~LeRose,$^{20}$ 
S.~Malov,$^{17}$ D.J.~Margaziotis,$^{2}$ J.~W.~Martin,$^{12}$ 
K.~McCormick,$^{12}$
R.~D.~McKeown,$^{1}$ K.~McIlhany,$^{12}$ Z.-E.~Meziani,$^{19}$
R.~Michaels,$^{20}$ G.~W.~Miller,$^{15}$
E.~Pace,$^{7,23}$ T.~Pavlin,$^{1}$ G.~G.~Petratos,$^{10}$ 
R.~I.~Pomatsalyuk,$^{9}$
D.~Pripstein,$^{1}$ D.~Prout,$^{10}$ R.~D.~Ransome,$^{17}$ Y.~Roblin,$^{11}$ 
M.~Rvachev,$^{12}$ A.~Saha,$^{20}$ G.~Salm\`{e},$^{6}$ M.~Schnee,$^{19}$ 
T.~Shin,$^{12}$ 
K.~Slifer,$^{19}$ P.~A.~Souder,$^{18}$ S.~Strauch,$^{17}$ R.~Suleiman,$^{10}$ 
M.~Sutter,$^{12}$ B.~Tipton,$^{12}$ L.~Todor,$^{14}$ M.~Viviani,$^{5}$ 
B.~Vlahovic,$^{13,20}$ 
J.~Watson,$^{10}$ C.~F.~Williamson,$^{10}$ H.~Wita{\l}a,$^{8}$ 
B.~Wojtsekhowski,$^{20}$ J.~Yeh,$^{18}$ P.~\.{Z}o{\l}nierczuk$^{21}$}

\address{$^{1}$California Institute of Technology, Pasadena, CA 91125, USA\\
$^{2}$California State University of Los Angles, Los Angles, CA 90032, USA\\
$^{3}$College of William and Mary, Williamsburg, VA~23187, USA\\
$^{4}$Duke University, Durham, NC~27708, USA\\
$^{5}$INFN, Sezione di Pisa, 56010 S.Piero a Grado, Pisa, Italy \\
$^{6}$INFN,Sezione di Roma, P.le A.  Moro 2, I-00185 Roma, Italy \\
$^{7}$INFN, Sezione Tor Vergata, Via della Ricerca Scientifica 1, 
I-00133 Rome, Italy \\
$^{8}$Institute of Physics, Jagellonian University, PL-30059 Cracow, Poland\\ 
$^{9}$Kharkov Institute of Physics and Technology, Kharkov 310108, Ukraine\\
$^{10}$Kent State University, Kent, OH~44242, USA\\
$^{11}$LPC, Universit\'{e} Blaise Pascal, F-63177 Aubi\`{e}re, France\\
$^{12}$Massachusetts Institute of Technology, Cambridge, MA~02139, USA\\
$^{13}$North Carolina Central University, Durham, NC~27707, USA\\
$^{14}$Old Dominion University, Norfolk, VA~23508, USA\\
$^{15}$Princeton University, Princeton, NJ~08544, USA\\
$^{16}$Ruhr-University, D-44780 Bochum, Germany\\
$^{17}$Rutgers University, Piscataway, NJ~08855, USA\\
$^{18}$Syracuse University, Syracuse, NY~13244, USA\\
$^{19}$Temple University, Philadelphia, PA~19122, USA\\
$^{20}$Thomas Jefferson National Accelerator Facility, Newport News, VA 23606,
 USA\\
$^{21}$University of Kentucky, Lexington, KY~40506, USA\\
$^{22}$University of New Hampshire, Durham, NH~03824, USA\\
$^{23}$Dipartimento di Fisica, Universit\`a di Roma "Tor Vergata", Rome, 
Italy\\
$^{24}$University of Virginia, Charlottesville, VA~22903, USA}

\date{01 July 2001}

\maketitle

\begin{abstract} 
We present the first precision measurement of the spin-dependent asymmetry in 
the threshold region of $^3\vec{\rm He}(\vec{e},e')$ at $Q^2$-values of 0.1 
and 0.2 (GeV/c)$^2$. The agreement between the data and non-relativistic 
Faddeev calculations which include both final-state interactions (FSI) and 
meson-exchange currents (MEC) effects is very good at $Q^2$ = 0.1 
(GeV/c)$^2$, while a small discrepancy at $Q^2$ = 0.2 (GeV/c)$^2$ is 
observed.
\end{abstract} 
 

        
Three-nucleon systems have been an excellent testing ground between theory 
and experiment in nuclear physics~\cite{glockle1}. In the context of 
electromagnetic processes, exact non-relativistic Faddeev 
calculations for both the ground state and the continuum of $^3$H and $^3$He 
have been carried out using a variety of modern nucleon-nucleon (NN) 
potentials~\cite{anklin1,golak1,golak2}. The exact treatment of final-state 
interactions (FSI) in the Faddeev calculation results in a much improved 
description of unpolarized $pd$ capture and breakup 
channels~\cite{anklin1,golak2}, 
as well as unpolarized electron scattering from the three-nucleon 
system~\cite{golak1}. This has provided important information on the nuclear 
ground-state structure and thus allows a deeper understanding of
the underlying nuclear force. With the availability of polarized beams and 
polarized targets, it has become possible to study additional spin-dependent 
quantities. Polarized $^3$He is an ideal target for such a study.
                  
Polarized $^3$He is also important as an effective neutron 
target~\cite{blankleider1,friar1}, because its ground state wave function is 
dominated by the $S$-state in which the proton spins cancel and the nuclear 
spin is carried entirely by the neutron. The spin-dependent asymmetries are 
thus sensitive to the neutron electromagnetic form factors in the vicinity of
the quasielastic peak of polarized electrons scattering from a polarized 
$^3$He 
target~\cite{blankleider1,friar1,ciofi1,schulze1,ciofi2,ishikawa1}. Recently 
there has been significant progress in extracting neutron electromagnetic form
factors from double-polarization electron-$^3$He scattering 
experiments~\cite{kotlyer1,xu1,meyerhoff1,becker1,rohe1}. In recent years, 
there have also been extensive 
efforts~\cite{anthony1,abe1,ackerstaff1,meziani1} in studying polarized 
inelastic 
scattering of electrons from polarized $^3$He targets in the deep inelastic 
and resonance regions aiming at understanding the underlying neutron spin 
structure. The extraction of the neutron spin structure functions from these 
experiments requires detailed knowledge of the $^3$He nuclear ground-state 
structure~\cite{friar1,ciofi3,ciofi4}. 
        
However, to probe the nuclear ground state structure, to extract 
the neutron electromagnetic form factors or to extract the neutron spin 
structure function in the resonance region, the reaction mechanism, especially
FSI and meson-exchange currents (MEC) effects, must be
well understood. Recently, a non-relativistic Faddeev calculation which
includes both FSI and MEC has been carried out \cite{golak2} for the first 
time, and describes very well the recent precision data ~\cite{xu1} on the 
transverse asymmetry $A_{T'}$ near the top of the quasielastic peak from the 
$^3\vec{\mathrm{He}}(\vec{e},e')$ process at low $Q^2$. However, since  
FSI and MEC effects are relatively small in this region, it 
is highly desirable to study another region where these two effects are 
larger 
to provide a more stringent constraint on the theory. The threshold region of 
$^3\vec{\mathrm{He}}(\vec{e},e')$, which extends from the two-body 
breakup threshold (with breakup energy of 5.5 MeV), the three-body breakup
threshold (with breakup energy of 7.7 MeV) to the low energy transfer side 
of the quasielastic peak, is an ideal place for such a study. First, FSI 
effects are expected to be large in the threshold region since the final 
state 
nucleons have less kinetic energy and thus have a higher probability of 
interacting with each other. Secondly, it has been shown that a substantial 
contribution from MEC is needed to describe the measured elastic 
electromagnetic form factors of the three-body system~\cite{amroun1}. 
Therefore one would expect a large MEC effect in the threshold region as 
well. 

A precision measurement of spin observables in the threshold 
region of $^3\vec{\mathrm{He}}(\vec{e},e')$ would thus provide us with
important information on the reaction mechanism, thereby 
placing significant constraints on the theoretical uncertainties in probing 
the $^3$He ground state structure and in extracting the neutron 
electromagnetic form factors from electron
scattering from $^3$He. In this Letter we report the first precision 
measurement of the spin-dependent asymmetry in the threshold region of 
$^3\vec{\mathrm{He}}(\vec{e},e')$.


For inclusive scattering of longitudinally polarized electrons from a 
polarized spin-1/2 target such as $^3$He, the spin-dependent asymmetry is 
defined as
$A = \frac{\sigma^{h+}-\sigma^{h-}}{\sigma^{h+}+\sigma^{h-}}$, 
where $\sigma^{h^{\pm}}$ are the cross sections for the two different 
helicities of the polarized electrons. It is given in terms of 
the quasielastic response functions as~\cite{donnelly1}
\begin{equation}
\label{asym}
A = \frac{-(\cos{\theta^{*}}\nu_{T'}R_{T'} +
  2\sin{\theta^{*}}\cos{\phi^{*}}\nu_{TL'}R_{TL'})}{\nu_{L}R_{L} +
  \nu_{T}R_{T}}
\end{equation}  
where the $\nu_k$ are kinematic factors and $\theta^{*}$ and
$\phi^{*}$ are the polar and azimuthal angles of target spin with
respect to the 3-momentum transfer vector ${\bf q}$ in the laboratory frame. 
$R_L$ and $R_T$ are the spin-independent longitudinal and transverse response 
functions, while $R_{T'}$ and $R_{TL'}$ are the spin-dependent transverse and 
longitudinal-transverse ones. The response functions depend on the electron 
energy transfer $\omega$ and the four-momentum transfer squared $Q^2$. By 
choosing $\theta^{*}$ = $0^{\circ}$ ($90^{\circ}$), one selects the transverse 
asymmetry $A_{T'}$ (longitudinal-transverse asymmetry $A_{TL'}$). 


The experiment was carried out in Hall A at the Thomas Jefferson National
Accelerator Facility (JLab), using a longitudinally polarized continuous wave 
electron beam of 10 $\mu$A current incident on a high-pressure polarized 
$^3$He gas target.  A detailed description of this experiment can be found in 
a previous publication~\cite{xu1}.

Electrons scattered from the target were detected in the two Hall A high
resolution spectrometers, HRSe and HRSh. The data from HRSe have been 
presented in a previous Letter~\cite{xu1}. The data from HRSh were used for 
this analysis and covered both the elastic peak and the threshold region. 
Since the elastic asymmetry can be calculated accurately at low $Q^2$ using 
the well-known elastic form factors of $^3$He~\cite{amroun2}, the elastic 
measurement allows a precise monitoring of the product of the beam and target 
polarizations, $P_bP_t$. Two kinematic points were measured in the threshold 
region, one with a central $Q^2$-value of 0.1 (GeV/c)$^2$ at an incident beam 
energy $E_0$ = 0.778 GeV and the other with a central $Q^2$-value of 0.2 
(GeV/c)$^2$ at $E_0$ = 1.727 GeV. The target spin was oriented at 
62.5$^\circ$ to the right of the incident electron momentum direction, while 
the outgoing electron momentum directions were 23.7$^\circ$ and 15.0$^\circ$ 
to the right of the incident electron momentum direction for $Q^2$ = 0.1 and 
0.2 (GeV/c)$^2$, respectively. This corresponds to $\theta^{*}$ from 
131.2$^\circ$ to 136.5$^\circ$ for $Q^2$ = 0.1 (GeV/c)$^2$, and 
from 134.2$^\circ$ to 140.0$^\circ$ for $Q^2$ = 0.2 (GeV/c)$^2$.


The yield for each electron helicity state was corrected by its corresponding
charge and computer dead time, and the raw experimental asymmetry was formed
as a function of the excitation energy in the $^3$He system, which is defined 
as $E_{x} = \sqrt{M^2 + 2M\omega - Q^2} - M$, where M is the mass of the 
$^3$He target. A 5 MeV bin was used for the excitation energy. The range 
of the excitation energy is from 5.5 MeV, which 
corresponds to the two-body breakup threshold, to about 35 MeV for $Q^2$ = 0.1
(GeV/c)$^2$, and about 50 MeV for $Q^2$ = 0.2 
(GeV/c)$^2$. The raw asymmetry was then corrected for dilutions due to 
scattering from the target walls, the nitrogen admixture inside the target
cell, and $P_bP_t$. 
The physics asymmetry was obtained after subtraction of the elastic 
radiative tail contribution, radiative correction of the quasielastic 
asymmetry, and correction for spectrometer acceptance and bin-averaging 
effects, all obtained from a Monte Carlo simulation~\cite{xiong1}. The
external radiative correction was treated following the standard procedure 
of Mo and Tsai~\cite{mo1}. The internal radiative correction was calculated 
using the covariant formalism of Akushevich et al.~\cite{aku1}. This
procedure requires knowledge of $^3$He nuclear response functions at various 
kinematics points, which were obtained from full Faddeev 
calculations~\cite{golak2}. 

\begin{figure}[t]
\psfig{file=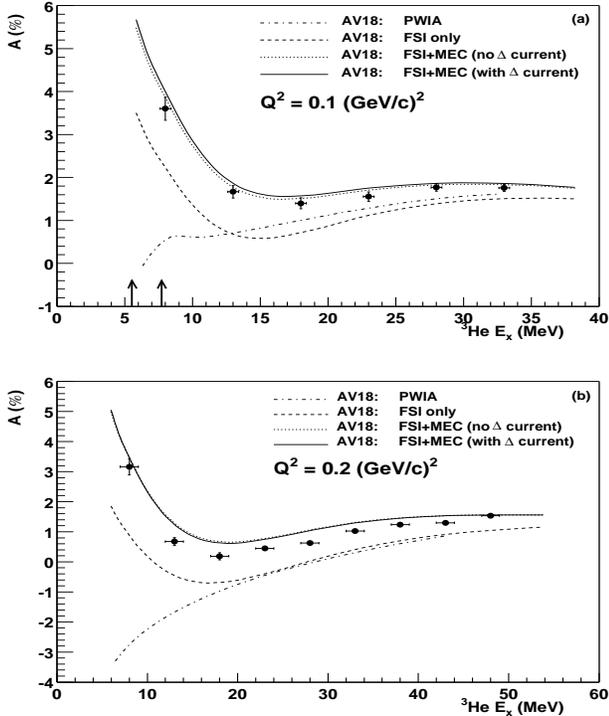,height=10.cm,width=9.cm}
\caption[]{The physics asymmetry together with theoretical calculations for 
     (a) $Q^2$ = 0.1 (GeV/c)$^2$
 and (b) $Q^2$ = 0.2 (GeV/c)$^2$. The theoretical calculations are all 
performed using AV 18 potential, but with different reaction mechanisms.
The arrows point to the two-body and three-body breakup thresholds.} 
\label{asy}
\end{figure}


Results for the physics asymmetry at both kinematics are shown in 
Fig.~\ref{asy}. The uncertainty in determining the excitation energy is
about 0.4 MeV at $Q^2$ = 0.1 (GeV/c)$^2$, and 1.0 MeV at $Q^2$ = 0.2 
(GeV/c)$^2$, dominated by the uncertainty in the beam energy. The vertical 
error bars on the data are the statistical and systematic errors added in 
quadrature. The systematic uncertainty includes contributions from the 
determination of $P_bP_t$, target wall and $N_2$ background 
subtraction, elastic radiative tail subtraction, radiative correction and the 
correction of spectrometer acceptance and bin-averaging effects. A careful 
analysis of systematic uncertainties was carried out and the results are shown
together with the physics asymmetry and statistical uncertainties in 
Table~\ref{sys1} for $Q^2$ = 0.1 (GeV/c)$^2$ and Table~\ref{sys2} for 
$Q^2$ = 0.2 (GeV/c)$^2$. 

\begin{table}
\begin{tabular}{c c c c c c c} 
 $E_x$ &$A \pm \delta^{stat}$&$\delta^{pol}$&$\delta^{dil}$ & $\delta^{ert}$  
            & $\delta^{rc}$  & $\delta^{acc}$ \\ 
(MeV)  &(\%)&   (\%)         &     (\%)     &     (\%)       
            &   (\%)         &     (\%)      \\ \hline  
$\ \:8.0$ & 3.602 $\pm$ 0.157 & 0.153 & 0.048 & 0.105 & 0.020 & 0.032 \\ 
13.0      & 1.666 $\pm$ 0.100 & 0.073 & 0.021 & 0.061 & 0.014 & 0.015 \\
18.0      & 1.399 $\pm$ 0.082 & 0.050 & 0.012 & 0.076 & 0.010 & 0.009 \\
23.0      & 1.553 $\pm$ 0.071 & 0.043 & 0.009 & 0.066 & 0.008 & 0.023 \\
28.0      & 1.768 $\pm$ 0.063 & 0.043 & 0.008 & 0.042 & 0.008 & 0.009 \\
33.0      & 1.756 $\pm$ 0.066 & 0.039 & 0.007 & 0.016 & 0.010 & 0.009 \\
\end{tabular}
\caption[]{Systematic uncertainties at each excitation energy ($E_x$) for 
$Q^2$ = 0.1 (GeV/c)$^2$, which include contributions from the determination 
of $P_bP_t$ ($\delta^{pol}$), target wall and $N_2$ dilution ($\delta^{dil}$),
elastic radiative tail subtraction ($\delta^{ert}$), radiative correction 
($\delta^{rc}$) and the correction of spectrometer acceptance and 
bin-averaging effects ($\delta^{acc}$). The physics asymmetry ($A$) and 
statistical uncertainties ($\delta^{stat}$) are also shown.} 
\label{sys1}
\end{table}

\begin{table}
\begin{tabular}{c c c c c c c} 
 $E_x$ & $A\pm \delta^{stat}$&$\delta^{pol}$ & $\delta^{dil}$ & $\delta^{ert}$ 
            & $\delta^{rc}$  & $\delta^{acc}$ \\  
(MeV)& (\%) &   (\%)         &     (\%)       &     (\%)       
            &   (\%)         &     (\%)      \\ \hline
$\ \:8.0$ & 3.161 $\pm$ 0.170 & 0.121 & 0.070 & 0.121 & 0.018 & 0.014 \\ 
13.0      & 0.676 $\pm$ 0.094 & 0.044 & 0.022 & 0.064 & 0.034 & 0.011 \\
18.0      & 0.190 $\pm$ 0.071 & 0.022 & 0.010 & 0.036 & 0.075 & 0.035 \\
23.0      & 0.446 $\pm$ 0.058 & 0.020 & 0.008 & 0.021 & 0.012 & 0.021 \\
28.0      & 0.625 $\pm$ 0.049 & 0.019 & 0.006 & 0.012 & 0.006 & 0.021 \\
33.0      & 1.025 $\pm$ 0.045 & 0.024 & 0.007 & 0.007 & 0.012 & 0.017 \\
38.0      & 1.241 $\pm$ 0.041 & 0.026 & 0.007 & 0.005 & 0.015 & 0.013 \\
43.0      & 1.300 $\pm$ 0.041 & 0.026 & 0.006 & 0.005 & 0.018 & 0.011 \\
48.0      & 1.537 $\pm$ 0.050 & 0.028 & 0.005 & 0.005 & 0.020 & 0.022 \\
\end{tabular}                     
\caption[]{Systematic uncertainties for $Q^2$ = 0.2 (GeV/c)$^2$. Symbols are
the same as in Table~\ref{sys1}.} 
\label{sys2}                      
\end{table}                       
                                   
All theoretical calculations were performed using AV18~\cite{wiringa1} as the 
NN interaction potential and the H\"{o}hler nucleon form factor 
parametrization\cite{hoh1}. Plane wave impulse approximation (PWIA) 
calculations~\cite{ciofi2,kievsky1} are shown as dot-dashed lines. 
Non-relativistic Faddeev calculations with FSI only~\cite{golak3} are shown 
as dashed lines. Non-relativistic Faddeev calculations which include both FSI 
and MEC~\cite{golak3} are shown as dotted lines without the inclusion of the 
$\Delta$ isobar current, and solid lines with the inclusion of the $\Delta$ 
isobar current. The MEC's ($\pi$ and $\rho$ exchanges) were chosen according
to a prescription given by Riska~\cite{riska1}, which guarantees to a large
extent the consistency of the MEC's to the NN force used. The agreement 
between the full calculation and the data is very good at $Q^2$ = 0.1 
(GeV/c)$^2$, and a relatively small discrepancy is observed at 
$Q^2$ = 0.2 (GeV/c)$^2$.

\begin{figure}[t]
\psfig{file=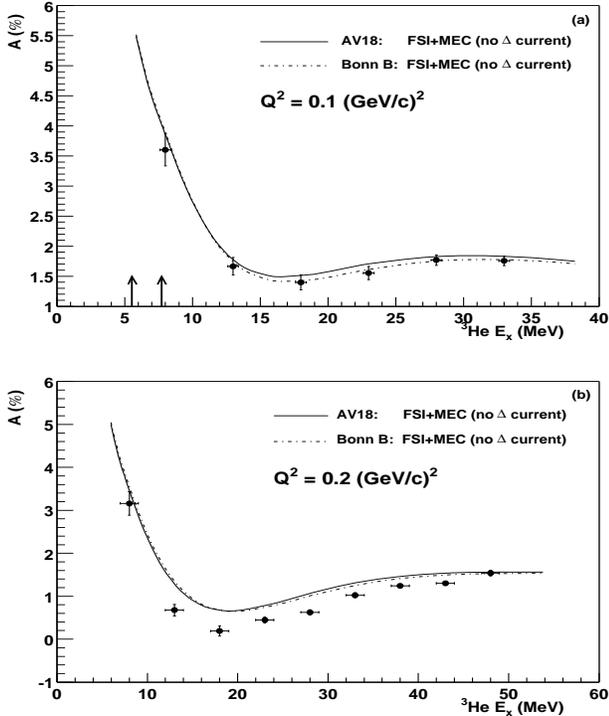,height=10.cm,width=9.cm}
\caption[]{The physics asymmetry together with two full Faddeev calculations, 
one using AV18 as the NN potential, the other using Bonn-B: (a) $Q^2$ = 0.1 
(GeV/c)$^2$ and (b) $Q^2$ = 0.2 (GeV/c)$^2$. } 
\label{tensor}
\end{figure}

To investigate the effects of different NN potentials, we compare our data 
with full Faddeev calculations using the AV18 potential and the Bonn-B 
potential~\cite{machleidt3}, a non-local potential which is very different
from the local AV18 potential. The result is shown in 
Fig.~\ref{tensor}. As can be seen, the difference between the theoretical
calculations using these two potentials is very small, which suggests that
this observable is not sensitive to the choice of different NN potentials 
and the corresponding exchange currents.

Theoretical uncertainties due to $G_E^p$, $G_M^p$,
$G_E^n$ and $G_M^n$ were studied using PWIA~\cite{friar1,ole1}. The relative
difference between the asymmetries calculated with the nucleon form factors 
from the H\"{o}hler parametrization and from 
data~\cite{xu1,meyerhoff1,becker1,rohe1,eden1,ostrick1,herberg1,passchier1} 
was found to be around 1\%, and thus theoretical uncertainties due to nucleon 
form factors are completely negligible. 

The good agreement between the full calculation and the data at $Q^2$ = 
0.1 (GeV/c)$^2$ suggests the validity of the current way of 
treating FSI and MEC in the full calculation. The small discrepancy at 
$Q^2$ = 0.2 (GeV/c)$^2$ may be due to the fact that some $Q^2$-dependent 
effects, such as the relativisitic effect, are not included 
in the current non-relativistic Faddeev calculation.   


In conclusion we have presented the first precision data on the 
spin-dependent 
asymmetry in the threshold region of $^3\vec{\mathrm{He}}(\vec{e},e')$. 
The agreement between the data and non-relativistic Faddeev calculations 
which include both FSI and MEC effects is very good at $Q^2$ = 
0.1 (GeV/c)$^2$, while the discrepancy at $Q^2$ = 0.2 (GeV/c)$^2$ might be 
due to some $Q^2$-dependent mechanism.


We thank the Hall A technical staff and the Jefferson Lab
Accelerator Division for their outstanding support during this experiment.
We also thank T.~W.~Donnelly for many helpful discussions.
This work was supported in part by the U.~S.~Department of Energy, DOE/EPSCoR,
the U.~S.~National Science Foundation, 
the Science and Technology Cooperation
Germany-Poland and the Polish Committee for Scientific Research, 
the Ministero dell'Universit\`{a} e della Ricerca
Scientifica e Tecnologica (Murst),
the French Commissariat \`{a} l'\'{E}nergie Atomique,
Centre National de la Recherche Scientifique (CNRS), Conseil R\'egional
d'Auvergne, the Italian Istituto Nazionale di Fisica
Nucleare (INFN), and grant of European Foundation Project INTAS-99-0125.
This work was supported by DOE contract DE-AC05-84ER40150
under which the Southeastern Universities Research Association
(SURA) operates the Thomas Jefferson National Accelerator Facility.
The numerical calculations were performed on the PVP machines at the U.~S. 
National Energy Research Scientific Computer Center (NERSC) and
the CRAY T90 of the NIC in
J\"{u}lich.


\end{document}